\documentclass[12pt,preprint]{aastex}
\begin{document}
\title{Have the Elusive Progenitors of Supernovae Type Ia Been Discovered?}
\author{Mario Livio and Adam Riess}
\affil{Space Telescope Science Institute\\
3700 San Martin Drive\\
Baltimore, MD 21218\\
email: mlivio@stsci.edu; ariess@stsci.edu}

\begin{abstract}
The recent detection of H$\alpha$ emission in the supernova Type~Ia SN~2002ic could be taken to mean that the elusive progenitor systems of Type~Ia supernovae have finally been identified.  At first glance, the observation appears to support a single-degenerate scenario, in which the white dwarf accretes from a normal companion.  In this Letter we show that the opposite may be true, and the observations may support the merger of two white dwarfs as the cause for Type~Ia supernovae.
\end{abstract}
\keywords{cosmology: observations -- supernovae: general}

\section{Introduction}
The recent detection of H$\alpha$ emission in the spectrum of the supernova Type~Ia (SN~Ia) SN~2002ic (Hamuy et~al.\ 2003) is a landmark discovery.  While there is very little doubt that SNe~Ia represent the thermonuclear disruption of mass accreting white dwarfs (WDs), the precise nature of the progenitor systems remains uncertain (Branch et~al.\ 1995; Livio 2001). Given that SNe~Ia are the tool of choice for confirming the acceleration of cosmic expansion (Riess et~al.\ 1998; Perlmutter et~al.\ 1999), the importance of identifying the progenitors cannot be overemphasized. The two main scenarios that have been proposed involve either the merger of two white dwarfs (the double-degenerate scenario; Iben \& Tutukov 1984; Webbink 1984), or a single white dwarf accreting from a normal companion (the single-degenerate scenario; Whelan \& Iben 1973; Nomoto 1982). Recently it has been argued theoretically, that single-degenerate progenitors are favored (even though it is very difficult for hydrogen-accreting WDs to reach the Chandrasekhar limit; Piersanti et al.\ 2000), and that double WD mergers may lead to accretion-induced collapses rather than to SNe~Ia (Livio 2001; Nomoto et~al.\ 2000).  The tentative discovery (if confirmed) of an enhanced SN~Ia rate near jets in active galactic nuclei (Livio, Riess, \& Sparks 2002; Capetti 2002) appears to support this conclusion. Nevertheless, until SN~2002ic the ``smoking gun''---the presence of hydrogen in the spectrum---was missing. The clear detection of a broad (FWHM~$\sim1800$~km~s$^{-1}$) H$\alpha$ component in SN~2002ic appears on the face of it to demonstrate that at least some SNe~Ia result from single-degenerate progenitors. In the present letter we show that this conclusion may be \emph{premature}.

\section{Why Now?}
One of the key questions posed by the observations of Hamuy et~al.\ (2003) is: Why was hydrogen not detected before? This becomes particularly puzzling when we realize that there exist about 100 spectra of SNe~Ia in which a signature of the strength of that seen in SN~2002ic would have been detected (T.~Matheson, private communication), had it been there. In fact, Hamuy et~al.\ noted that the amount of shock-heated circumstellar material needed to produce the observations of SN~2002ic is totally unexpected for a SN~Ia.  Accordingly, they suggested that the progenitor system was a binary consisting of a C/O white dwarf and a \emph{massive} (3--7~M$_{\odot}$) asymptotic giant branch (AGB) star. The presence of the latter was necessitated by the need to have an integrated circumstellar mass of at least a few solar masses.

The main problem with this scenario is that one would expect to observe \emph{a range} of strengths of H$\alpha$ lines in SNe~Ia, depending on the amount of circumstellar material (in turn, determined primarily by the mass of the AGB star), rather than \emph{detecting a relatively strong line in one case only} (it is also hard to believe that this is the first progenitor system containing an AGB star). 

We propose instead that the total absence of H$\alpha$ lines in all the pre-SN~2002ic SNe~Ia observed to date argues that SN~2002ic represents rather rare circumstances, and \emph{not} a white dwarf accreting from the wind of an AGB star.

\section{A Supernova Ia in a Common Envelope?}
All the evolutionary scenarios leading to the formation of close double white dwarf systems involve a stage in which an AGB star fills its Roche lobe and transfers mass onto a white dwarf companion (e.g.\ Yungelson \& Livio 2000). Under these conditions, the mass transfer process is unstable, and the system evolves rapidly into a common envelope (CE) configuration, inside which the white dwarf and the core of the AGB star spiral-in (e.g.\ Rasio \& Livio 1996; Taam \& Sandquist 2000). Typically, the CE phase lasts a few hundred to a few thousand years, and results in the ejection of the envelope and the emergence of a double white dwarf system (e.g.\ Sandquist et~al.\ 1998; Taam \& Sandquist 2000 and references therein). I propose that SN~2002ic represents one of those rare cases in which the explosion occurs \emph{during} (or immediately following) the CE phase, and in which some part of the envelope has not been previously ejected. This raises two immediate questions: (i)~Is this possible at all? and (ii)~Does this support a single-degenerate or a double-degenerate scenario?

For the white dwarf to actually reach the Chandrasekhar mass via accretion of hydrogen-rich material during the CE phase is extraordinarily unlikely. Steady burning occurs for a narrow range of accretion rates of order (Paczy\'nski \& \.Zytkow 1978; Nomoto, Nariai \& Sugimoto 1979; the limits are determined:  at the low end by the requirement that the pressure at the time of ignition be sufficiently low to prevent a shell flash, and at the high end by the accretor expanding to supergiant dimensions)
\begin{equation}
0.4\ \dot{M}_\mathrm{RG}\lesssim\dot{M}\lesssim\dot{M}_\mathrm{RG}~~.
\end{equation}
Here $\dot{M}_\mathrm{RG}$ is the rate at which the white dwarf expands to giant dimensions and is given by
\begin{equation}
\dot{M}_\mathrm{RG}\simeq8.5\times10^{-7} (M_\mathrm{WD}/\mathrm{M}_{\odot}-0.52)\ \mathrm{M}_{\odot}/\mathrm{yr}~~.
\end{equation}
Even assuming that the accretion rate could be regulated to the rate given by equation~(1) [most likely it would settle on the Eddington rate of $\dot{M}_\mathrm{EDD}\simeq1.7\times10^{-5} (R_\mathrm{WD}/10^{9}~\mathrm{cm})\ \mathrm{M}_{\odot}/\mathrm{yr}$ at which mass would not be retained], the WD would increase in mass by at most $\sim$0.001~M$_{\odot}$ during the CE phase. This would require the WD to be within 0.001~M$_{\odot}$ of the Chandrasekhar mass upon entering the CE---a very unlikely situation, even taking into account the rarity of H$\alpha$ detection (e.g.\ only 2 out of a sample of 130 WDs were found to have masses higher than 1.2~M$_{\odot}$; Bergeron, Saffer, \& Liebert 1992; although see Hachisu \& Kato 1999).

A second possibility is that the WD spirals-in all the way to the center and merges with the AGB star's core. Interestingly, a scenario for SNe of similar type was suggested almost 30 years ago by Sparks \& Stecher (1974), but has long since been discarded due to the absence of hydrogen in the spectra. What I propose here is that the spiraling-in process unbinds most, but not all of the envelope, so that coalescence becomes inevitable. At the time of merger, most of the envelope will be at a distance of 
\begin{equation}
d\simeq3\times10^{15}
\left(\frac{V}{10~\mathrm{km~s}^{-1}}\right)
\left(\frac{\tau_\mathrm{CE}}{100~\mathrm{yr}}\right)\ \mathrm{cm}~~.
\end{equation}
from the core.  Here $V$ is the ejection velocity and $\tau_\mathrm{CE}$ is the duration of the CE phase. The condition for a merger to occur (as opposed to ejection of the entire envelope and the formation of a binary WD system) is given by the requirement that the binding energy of the CE be larger than the gravitational energy available from orbital shrinkage (Livio 1996; deKool 1990)
\begin{equation}
\frac{M_\mathrm{AGB}(M_\mathrm{AGB}-M_C)}{\lambda\, a_0\, r_L}
>\alpha_\mathrm{CE}
\left(\frac{M_C M_\mathrm{WD}}{2R_C}-
\frac{M_\mathrm{AGB}M_\mathrm{WD}}{2a_0}\right)~~.
\end{equation}
Here $a_0$ is the initial separation, $r_L$ is the Roche lobe radius of the AGB star (in units of the separation), $M_C$ and $R_C$ are the mass and radius of the core, respectively, $\alpha_\mathrm{CE}$ is the CE efficiency parameter (Livio \& Soker 1988; Iben \& Tutukov 1984), and $\lambda\sim0.5$ depends on the stellar density profile. The value of $\alpha_\mathrm{CE}$ is not known even to within a factor 10 (e.g.\ Livio 1996). However, for reasonable values ($\alpha_\mathrm{CE}\sim0.1$--1) condition~(4) requires relatively massive AGB stars [since the condition can be approximated as $(M_\mathrm{AGB}/M_\mathrm{WD})^2\gtrsim1/8\,\alpha_\mathrm{CE}(a_0/R_C)$; and $a_0/R_C\sim10^4$] and can be expected to be satisfied only in a fraction of a percent of all systems (e.g.\ Yungelson \& Livio 1998). The observed H$\alpha$ emission would result from the interaction of the explosion with the previously-ejected envelope. This would be consistent with the rarity of H$\alpha$ detections. Most importantly, however, if this scenario is correct, the H$\alpha$ detection by Hamuy et~al.\ \emph{results from a double-degenerate scenario}!

\section{Conclusions}
One might have thought that the detection of hydrogen in the spectrum of a SN~Ia would have finally revealed the elusive progenitor to be a single-degenerate system. In this Letter we suggest that this may not be the case. Paradoxically, the H$\alpha$ detection could result from a double-degenerate scenario! To be sure, the actual result of the merger process remains as uncertain as ever, and it may lead to an accretion-induced collapse rather than to a SN~Ia. Other exotic possibilities, such as the explosion of the core of an AGB star (``type 1.5'' event; Iben \& Renzini 1983) may exist (as already suggested by Hamuy et~al.\ 2003). However, the latter would require some other mechanism to place (at least a part of) the envelope at $\sim10^{15}$~cm.  Future, more sensitive, observations will reveal whether the detection of H$\alpha$ is a very rare, but relatively clear event, or whether a range of line strengths is detected. The latter case would clearly support a single-degenerate interpretation.

\acknowledgements
We would like to thank David Branch and Tom Matheson for helpful discussions.

\end{document}